# Lattice dynamics and elastic properties of α-U at high-temperature and high-pressure by machine learning potential simulations

Hao Wang[1,2], Xiao-Long Pan[1,2], Yu-Feng Wang[2], Xiang-Rong Chen[1*], Yi-Xian Wang[3], Hua-Yun Geng[2,4*]

[1] *College of Physics, Sichuan University, Chengdu 610065, People's Republic of China*

[2] *National Key Laboratory of Shock Wave and Detonation Physics, Institute of Fluid Physics, CAEP, Mianyang 621900, People's Republic of China*

[3] *College of Science, Xi'an University of Science and Technology, Xi'an 710054, People's Republic of China*

[4] *HEDPS, Center for Applied Physics and Technology, and College of Engineering, Peking University, Beijing 100871, People's Republic of China*

**Abstract:** Studying the physical properties of materials under high pressure and temperature through experiments is difficult. Theoretical simulations can compensate for this deficiency. Currently, large-scale simulations using machine learning force fields are gaining popularity. As an important nuclear energy material, the evolution of the physical properties of uranium under extreme conditions is still unclear. Herein, we trained an accurate machine learning force field on α-U and predicted the lattice dynamics and elastic properties at high pressures and temperatures. The force field agrees well with the *ab initio* molecular dynamics (AIMD) and experimental results and it exhibits higher accuracy than classical potentials. Based on the high-temperature lattice dynamics study, we first present the temperature-pressure range in which the Kohn anomalous behavior of the $\Sigma_4$ optical mode exists. Phonon spectral function analysis showed that the phonon anharmonicity of α-U is very weak. We predict that the single-crystal elastic constants $C_{44}$, $C_{55}$, $C_{66}$, polycrystalline modulus ($E$, $G$), and polycrystalline sound velocity ($C_L$, $C_S$) have strong heating-induced softening. All the elastic moduli exhibited compression-induced hardening behavior. The Poisson's ratio shows that it is difficult to compress α-U at high pressures and temperatures. Moreover, we observed that the material becomes substantially more anisotropic at high pressures and temperatures. The accurate predictions of α-U

---

demonstrate the reliability of the method. This versatile method facilitates the study of other complex metallic materials.

## 1. Introduction

With the increasing shortage of energy today, metallic uranium, as an important nuclear energy material, has always attracted much attention[1]. Uranium exists stably in an orthogonal structure (α-U) at normal temperature and pressure[2]. In addition, it is the elemental substance that can be observed the behavior of charge density waves (CDW) at normal pressure.

Crummett *et al*. measured phonon dispersion behavior of α-U at room temperature by neutron inelastic scattering experiments, and found that the optical branch in the middle of the [100] direction has a huge softening behavior[3]. Smith *et al*. further experimentally confirmed that this softening behavior increased further with decreasing temperature, showing a soft-mode-driven phase transition[4]. Bouchet reproduced the phonon dispersion behavior of α-U by using the density functional perturbation theory (DFPT)[5]. In addition, they found that the pressure effect has a strong influence on the softening of the α-U optical branch. This pressure effect was then experimentally confirmed by Raymond *et al*[6]. Bouchet *et al*.[7, 8] further studied the phonon dispersion behavior of uranium at high temperature and high pressure by *ab initio* molecular dynamics (AIMD) calculation combined with the temperature-dependent effective potential technique (TDEP)[9, 10].

Both theory and experiments show that α-U has one CDW and multiple Kohn anomalies at $\mathbf{q}_{CDW}$. Roy *et al*.[11] showed that various Kohn anomalies in α-U arise from the combined effect of Fermi surface nesting (FSN) and "hidden" nesting, that is, the nesting of electronic states above and below the Fermi surface. The topology in favor of Fermi surface nesting (FSN) allows the electron susceptibility $\chi_0$ to diverge and induce a CDW at the wave vector $\mathbf{q}_{CDW}$.

At present, the mechanical study of uranium under high-temperature and high-pressure is lacking, especially for α-U. Fisher once experimentally measured the variation of the elastic constant of α-U with temperature at ambient pressure, and found that the $C_{11}$ of α-U had anomalous behavior at low temperature[12]. Due to the low symmetry of α-U, to measure the elastic constant at high pressure is much more difficult. Density functional theory (DFT) can accurately describe the change of the elastic constant of crystal at high pressure [13-15]. In order to take the temperature effect into account, however, AIMD simulation must be used. Unfortunately, due to the complexity of the α-U structure, the computational cost required to calculate its elastic constant by AIMD is much larger than that of materials with high structural symmetry such as Al[16] and TiN[17]. Bouchet *et al*.[8] gave a relationship between the bulk modulus and shear modulus of uranium with pressure and temperature through AIMD+TDEP. However, the variation of mechanical anisotropy and the sound velocity with pressure and temperature is unknown.

It should be pointed out that lattice parameters, phonon properties and elastic constants of α-U calculated by using the classical interatomic potential of uranium currently have large discrepancy against the experimental data[18-22]. In recent years, there have many different machine learning (ML) potential models been developed, such as Neural network potentials (NNP)[23], Gaussian approximation potentials (GAP)[24, 25], Moment tensor potentials (MTP)[26, 27], Deep learning potentials (DPMD)[28-30], etc. These models have a good performance for materials at high temperature and high pressure, such as Si[31], Fe[32], Ti[33, 34], Zr[35, 36], and alloy compounds[37, 38]. Existing results indicate that MTP and GAP have higher accuracy, while MTP is more efficient than GAP[39].

So far, investigation on the ML potential of uranium is still few. Although Kruglov *et al*.[40] studied uranium by the moment tensor potential method, they mainly focused on the phase transition, and did not report other physical properties. Ladygin *et al*.[41] mainly used γ-U to illustrate the accuracy of MTP in description of high-temperature phonon dispersion behavior.

In this paper, we train an accurate machine learning force field based on DFT data. With this model, we resolve the lattice dynamics and elastic properties of α-U at high temperature and high pressure. Machine learning force field successfully predicts the Kohn anomaly in the lattice dynamics of α-U at high temperature and high pressure. We calculate the elastic behavior of α-U, giving its dependence on pressure and temperature. Meanwhile, we characterize its elastic anisotropy under high temperature and high pressure.

## 2. Methods

### 2.1. Moment tensor potentials

In order to study the physical properties of α-U, we employ a machine learning model called moment tensor potentials (MTPs)[26]. The MTP training is performed with the software named "MLIP"[27]. MTP is a local potential. In this model, the total energy of the structure ($E^{\mathrm{mtp}}$) is expressed as the sum of atomic contributions $V(n_i)$, defined as

$$E^{\mathrm{MTP}}(\mathrm{cfg}) = \sum_{i=1}^{N} V(\boldsymbol{n}_i) \tag{1}$$

where $N$ is the total number of atoms in that configuration. The atomic environment neighborhood ($\boldsymbol{n}_i$) is composed of the $i$-th atomic type ($Z_i$), the nearest neighbor atomic type ($Z_j$) and the relative position ($\boldsymbol{r}_{ij}$) within a given cutoff radius.

In MTP, $V(\boldsymbol{n}_i)$ can be expressed as a linear combination of a set of basic functions ($B_\alpha(n_i)$), which is defined as follows

$$V(\boldsymbol{n}_i) = \sum_{\alpha} \xi_\alpha B_\alpha(\boldsymbol{n}_i) \tag{2}$$

In order to define the functional form of the basis $B_\alpha$, the form of moment tensor descriptors is introduced to describe the local information of atoms, as follows

$$M_{\mu,\nu}(\boldsymbol{n}_i) = \sum_{j} f_\mu(c) \underbrace{\boldsymbol{r}_{ij} \otimes \ldots \otimes \boldsymbol{r}_{ij}}_{\nu \text{ times}}; \tag{3}$$

where $f_\mu(c)$ and $\boldsymbol{r}_{ij} \otimes \ldots \otimes \boldsymbol{r}_{ij}$ represents the radial and angular part, respectively. The symbol $\otimes$ denotes the outer product of vectors. The $\mu, \nu \geq 0$ represents the different descriptors. The $\boldsymbol{\xi}$ and $\boldsymbol{c}$ constitute the parameters $\boldsymbol{\theta}(\boldsymbol{\xi}, \boldsymbol{c})$ we need to determine during the training process:

$$\sum_{k=1}^{K} \left[ w_e \left( E^{\mathrm{mtp}}(\mathrm{cfg}_k; \boldsymbol{\theta}) - E^{\mathrm{qm}}(\mathrm{cfg}_k) \right)^2 + w_f \sum_{i=1}^{N_k} \left| \mathbf{f}_i^{\mathrm{mtp}}(\mathrm{cfg}_k; \boldsymbol{\theta}) - \mathbf{f}_i^{\mathrm{qm}}(\mathrm{cfg}_k) \right|^2 \right. \\ \left. + w_s \left| \boldsymbol{\sigma}^{\mathrm{mtp}}(\mathrm{cfg}_k; \boldsymbol{\theta}) - \boldsymbol{\sigma}^{\mathrm{qm}}(\mathrm{cfg}_k) \right|^2 \right] \to \min_{\boldsymbol{\theta}} \tag{4}$$

where $N_k$ is the number of atoms in the $k$-th configuration, $E$, $\mathbf{f}$, $\boldsymbol{\sigma}$ represent energy, forces and virial stress, respectively, $w_e$, $w_f$, $w_s$ represent the non-negative weights of energy, force, and stress during training. For theoretical details of MTPs, such as the specific relationship between $B_\alpha$ and $M_{\mu,\nu}$, please refer to the Ref. [26, 27, 42]. The training settings can be found in Table I.

**Table I.** MTP training set size, training parameter settings.

| | |
|---|---|
| Number of structures | 3600 |
| Number of fitting parameters | 864 |
| Energy weight ($w_e$) | 1 |
| Force weight ($w_f$) | 0.1 |
| Stress weight ($w_s$) | 0.01 |
| Cut-off radius (Å) | 6.0 |

### 2.2. Elastic constants

In molecular dynamics, the elastic constants were calculated using the stress-strain method[43]. According to Hooke's law, the relationship between stress and strain tensor under Voigt symbol is defined as

$$\sigma_i = \sum_j C_{ij} \varepsilon_j \tag{5}$$

where $\sigma_i$ refers to stress tensor, $C_{ij}$ represents elastic constants, and $\varepsilon_j$ refers to strain tensor. We can obtain all the elastic constants of a material from Eq. (5), by applying the strain value and calculating the corresponding stresses

Because α-U belongs to orthorhombic system and has nine independent elastic constants. For obtaining $C_{11}$, $C_{12}$ and $C_{13}$, we use a strain tensor:

$$\varepsilon = \begin{bmatrix} \eta & 0 & 0 \\ 0 & 0 & 0 \\ 0 & 0 & 0 \end{bmatrix} \tag{6}$$

For $C_{22}$ and $C_{23}$, a strain tensor follows:

$$\varepsilon = \begin{bmatrix} 0 & 0 & 0 \\ 0 & \eta & 0 \\ 0 & 0 & 0 \end{bmatrix} \tag{7}$$

For $C_{33}$, $C_{44}$, $C_{55}$ and $C_{66}$, a strain tensor follows:

$$\varepsilon = \begin{bmatrix} 0 & \frac{\eta}{2} & \frac{\eta}{2} \\ \frac{\eta}{2} & 0 & \frac{\eta}{2} \\ \frac{\eta}{2} & \frac{\eta}{2} & \eta \end{bmatrix} \tag{8}$$

where $\eta$ represents the strain value. The deformed crystal lattice vector is

$$a' = a(I + \varepsilon) \tag{9}$$

where $a'$ and $a$ represent the crystal lattice vectors before and after deformation, respectively, and $I$ represent the identity matrix.

Then, the polycrystalline moduli are calculated by using the Voigt-Reuss-Hill (VRH) method[44, 45].

Bulk modulus ($B$):

$$\begin{aligned} 9B_V &= (C_{11} + C_{22} + C_{33}) + 2(C_{12} + C_{23} + C_{13}) \\ B_R &= \frac{1}{(S_{11} + S_{22} + S_{33}) + 2(S_{12} + S_{23} + S_{13})} \\ B_H &= \frac{B_V + B_R}{2} \end{aligned} \tag{10}$$

Shear modulus ($G$):

$$\begin{aligned} 15G_V &= (C_{11} + C_{22} + C_{33}) - (C_{12} + C_{23} + C_{13}) + 4(C_{44} + C_{55} + C_{66}) \\ G_R &= \frac{15}{4(S_{11} + S_{22} + S_{33}) - 4(S_{12} + S_{23} + S_{13}) + 3(S_{44} + S_{55} + S_{66})} \\ G_H &= \frac{G_V + G_R}{2} \end{aligned} \tag{11}$$

Young's modulus ($E$):

$$E_{V,R,H} = \frac{9B_{V,R,H} G_{V,R,H}}{3B_{V,R,H} + G_{V,R,H}} \tag{12}$$

where $S_{ij} = [C_{ij}]^{-1}$, called the compliance tensor. In which the quantity with subscript H (Hill) is what we reported in this work.

For isotropic and homogeneous solid, the longitudinal ($C_L$), shear ($C_S$) and bulk ($C_b$) sound velocities can be obtained as follows[46]:

$$C_L = \sqrt{\frac{B + \frac{4}{3}G}{\rho}}, \quad C_S = \sqrt{\frac{G}{\rho}}, \quad C_b = \sqrt{C_l^2 - \frac{4}{3}C_s^2} \qquad (13)$$

where $\rho$ is the density, and $B$ and $G$ represent the bulk and shear modulus of the solid, respectively.

Single crystals are basically anisotropic. In order to better measure the anisotropic characteristics of materials, Ranganathan *et al.*[47] proposed a universal anisotropy index including the contribution of the elastic stiffness tensor:

$$A^U = 5\frac{G_V}{G_R} + \frac{B_V}{B_R} - 6 \qquad (14)$$

When $A^U$ equal zero indicates that the material is locally isotropic. In addition, we can represent the anisotropic characteristics of materials through the spatial distribution of elastic modulus, such as Young's modulus[48], which is defined as follows

$$\frac{1}{E} = l_1^4 S_{11} + l_2^4 S_{22} + l_3^4 S_{33} + 2l_1^2 l_2^2 S_{12} + 2l_1^2 l_3^2 S_{13} + 2l_2^2 l_3^2 S_{23} + l_1^2 l_2^2 S_{66} + l_1^2 l_3^2 S_{55} + l_2^2 l_3^2 S_{44} \quad (15)$$

where $l_1$, $l_2$, and $l_3$ are the direction cosines.

## 2.3. Computational details

The DFT calculations were performed using the VASP package[49, 50]. A plane-wave basis set was employed with a kinetic energy cut-off of 500 eV. The electron-core interaction was described using a projector-augmented wave pseudopotential[51]. The electronic exchange-correlation functional was set to the generalized gradient approximation as the parameterized by Perdew, Berke, and Ernzerhof[52].

The required dataset was generated by the AIMD. The supercell size was 4 × 2 × 3, with 96 atoms. To obtain an atomic environment, simulations were performed using an isothermal–isobaric ensemble (NPT). We selected 0, 50, and 100 GPa. Four temperature points were selected for the

calculation under each pressure (0 GPa:100, 300, 600, 900 K; 50, 100 GPa:100, 300, 750, and 1200 K) with 12 NPT simulations. The AIMD simulation step was 2.5 fs and the total time was 3 ps. After the simulation was completed, we selected a training structure every 10 fs for 3600 structures.

The lattice dynamics and elastic constants were calculated using the TDEP and stress-strain methods, respectively. In AIMD, the size of the supercell was $5 \times 3 \times 3$ and $4 \times 2 \times 3$, containing 180 and 96 atoms, respectively. The canonical ensemble (NVT) was used to ensure invariance of volume and temperature in the AIMD calculations. The AIMD simulation step was 2.5 fs, and the total times were 5 ps (phonon) and 4 ps (elastic constant). The strain values $\eta$ used for the calculation of the elastic constant were -0.02, 0, and 0.02.

For all the AIMD calculations, the K-point sampling grid in the irreducible Brillouin zone was $3\times3\times2$. The energy and force convergence tolerances were set to $10^{-5}$ eV, respectively.

Molecular dynamics simulations with ML potentials were performed using the LAMMPS package[53]. The supercell size was $10 \times 4 \times 5$ and it contained 800 atoms. The step size of the molecular dynamics simulation was 1 fs, the total time was 50 ps, and the last 30 ps was used for averaging the thermodynamic quantities. To calculate the lattice constants of α-U at different temperatures and pressures, the *tri* relaxation in the NPT ensemble was utilized, and the pressure (temperature) damping parameters for the thermostat were set as 1.0 (0.1). Lattice dynamics and elastic constants were calculated using an NVT ensemble. The strain values $\eta$ used in the calculation of the elastic constants were -0.02, -0.01, 0, 0.01, and 0.02.

## 3. Results and discussion

Here, we discuss the application of MTP to simulate the properties of α-U at high temperatures and pressures.

### 3.1. Computational accuracy and efficiency of MTP

The fitting quality of the MTP must be checked before starting the study. Our training set contained 3600 structures with 345600 atomic environment information. The accuracy of the trained energy (force, stress) reaches a root-mean-square error (RMSE) of 9.37 meV (60.18 meV/Å, 0.98 GPa). As shown in Fig. 1, we further analyzed the differences between the energy, force, and stress predicted by MTP and AIMD at different pressures and temperatures. The RMSEs for energy, force, and stress were in the range of 1–17 meV/atom, 21–98 meV/Å, and 0.4–1.9 GPa, respectively. The energy training accuracy of our MTP is similar to that of Kruglov *et al*.[40] Notably, the MTP trained by Kruglov *et al*. is not publicly available for download, and the MTP trained by Ladygin *et al*.[41] cannot successfully simulate α-U in our tests. The results indicate that the force field accurately reflects DFT accuracy.

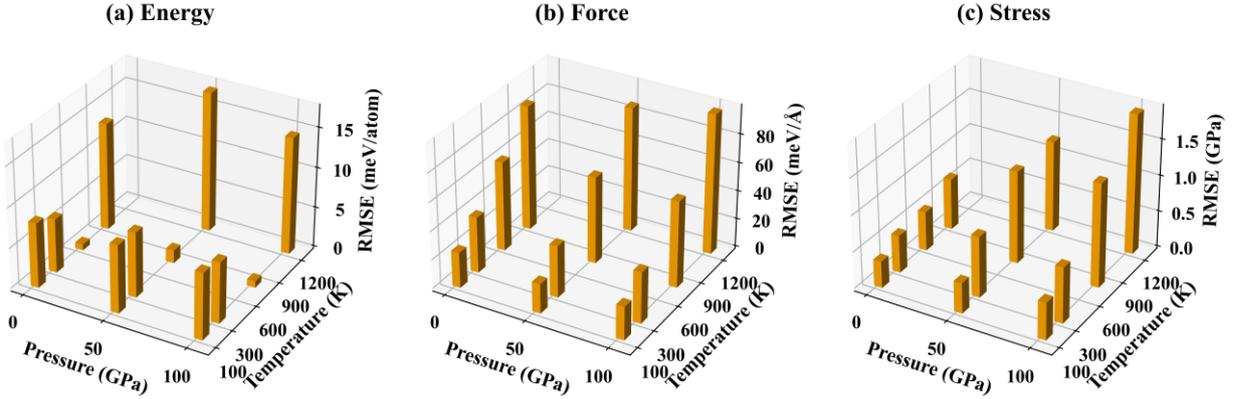

**Fig. 1**. (Color online) Root-mean-square errors of MTP at different pressures and temperatures. (a) Energy, (b) force, and (c) stress.

Table II lists the lattice parameters and elastic constants obtained using the different methods at 300 K and 0 K. The results calculated by MTP and AIMD were consistent with the experimental results[12, 54]. In comparison, the results obtained using existing classical potentials deviate significantly from the experimental results. For instance, the classical potential generally overestimates the bulk modulus and underestimates the shear modulus and Young's modulus more than MTP.

To investigate the reliability of the MTP for the prediction of the lattice dynamic properties of α-U further, we report the phonon spectrum in Fig. 2. The AIMD and MTP results are consistent. Although the theoretical results differ from the neutron scattering experiments[3] in the high-frequency optical branch, both the MTP and AIMD reproduce the Kohn anomaly well. In particular, the abnormal softening behavior of the $\Sigma_4$ branch in the middle of the [100] direction. The minimum values of the $\Sigma_4$ branch predicted by MTP and AIMD were located at **q** = [0.5 0 0] (0.99 THz) and **q**

= [0.5 0 0] (0.82 THz), respectively. The experimental result was **q** = [0.475 0 0] (1.01 THz)[3]. In contrast, the existing MEAM[18] and EAM[19] failed in terms of the lattice dynamics of α-U (Fig. 2(b)). A more detailed comparison of MTP reliability can be found in the Supplementary Material [55].

**Table II**. Comparison of lattice constants and elastic constants of α-U calculated by MTP, AIMD, and classical potentials[18-21] with experimental data[12, 54] at 300 K. Notably, MEAM, EAM1, and ADP results are from our recalculation, as well as EAM2 and COMB results refer to Ref. [18, 20, 22]. The FP-LMTO results[56] at 0 K are also summarized in table.

|  | MTP | AIMD | MEAM[18] | EAM1[19] | EAM2[20] | ADP[21] | COMB[22] | DFT[56] | Expt.[12, 54] |
|---|---|---|---|---|---|---|---|---|---|
| $a$ (Å) | 2.833 | 2.834 | 2.734 | 2.853 | 2.755 | 2.847 | 2.778 | 2.845 | 2.854 |
| $b$ (Å) | 5.828 | 5.832 | 6.361 | 5.740 | 6.072 | 5.842 | 6.152 | 5.818 | 5.869 |
| $c$ (Å) | 4.931 | 4.928 | 4.873 | 4.951 | 4.936 | 4.993 | 4.797 | 4.996 | 4.955 |
| $V$ (Å$^3$) | 20.36 | 20.36 | 21.19 | 20.27 | 20.79 | 20.76 | 20.49 | 20.67 | 20.75 |
| $C_{11}$ (GPa) | 211.73 | 195.68 | 382.9 | 137.3 | 182 | 237 | 257.6 | 300 | 215 |
| $C_{22}$ (GPa) | 206.18 | 209.76 | 197.5 | 205.1 | 138 | 198 | 222.6 | 220 | 199 |
| $C_{33}$ (GPa) | 302.47 | 318.27 | 241.8 | 303.3 | 162 | 249 | 298.9 | 320 | 267 |
| $C_{44}$ (GPa) | 93.79 | 131.58 | 74.7 | 64.7 | 26 | 65 | 100.0 | 150 | 124 |
| $C_{55}$ (GPa) | 89.95 | 93.15 | 39.9 | 60.5 | 21 | 36 | 61.7 | 93 | 73 |
| $C_{66}$ (GPa) | 97.38 | 82.81 | 68.1 | 38.3 | 21 | 48 | 89.2 | 120 | 74 |
| $C_{12}$ (GPa) | 51.65 | 37.88 | 75.7 | 98.3 | 105 | 91 | 99.1 | 50 | 46 |
| $C_{13}$ (GPa) | 40.85 | 37.24 | 27.1 | 116.7 | 121 | 87 | 46.0 | 5 | 22 |
| $C_{23}$ (GPa) | 124.75 | 138.57 | 87.0 | 105.7 | 103 | 104 | 66.5 | 110 | 108 |
| $B$ (GPa) | 124.81 | 121.25 | 131.4 | 134.5 | 125 | 138 | 133.5 | 129 | 113 |
| $G$ (GPa) | 87.33 | 91.35 | 72.1 | 50.9 | 24 | 54 | 85.4 | 112 | 84.4 |
| $E$ (GPa) | 212.44 | 219.05 | 182.7 | 135.6 | 67 | 144 | 211.2 | 262 | 202.6 |

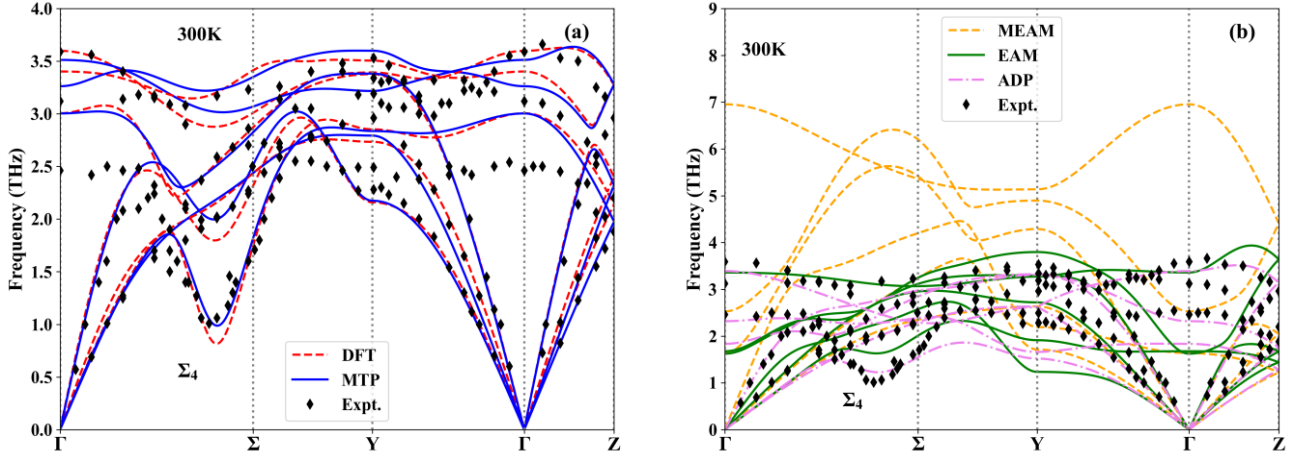

**Fig. 2**. (Color online) Phonon dispersion behavior of α-U at room temperature. (a) The results of MTP and AIMD; (b) The results of MEAM[18], EAM1[19], and ADP[21]. The solid black diamonds represent the experimental results[3].

We compared the computational efficiencies of the MTP, AIMD, EAM, MEAM, and ADP methods for the α-U study. Two Intel Xeon Silver 4114 processors with 20 cores and 40 threads (10 cores and 20 threads/processor) were used. For the AIMD calculations, a supercell containing 96 atoms was used for testing, K-point sampling was used a single Γ point, and the energy accuracy was sets to $10^{-4}$ eV. MTP and classical potential were tested using a supercell containing 800 atoms. The computational efficiency of MTP (0.058 s/step, 800 atoms) is significantly higher than that of AIMD (220 s/step, 96 atoms). Although MTP is 60–90 times slower than MEAM (0.0011 s/step) and EAM (0.0007 s/step), it has a higher accuracy. The MTP achieves a good balance between computational efficiency and computational accuracy.

### 3.2. Lattice dynamics at high temperatures and pressures

Based on the MTP, we systematically studied the lattice dynamics of α-U at high temperatures and pressures (Fig. 3). High pressure induces the stiffening of phonon frequencies. When the temperature increases, except for the $\Sigma_4$ branch at **q** = [0.5 0 0], the phonon frequency soften. Near the Γ point, the phonon thermal softening of the transverse acoustic modes $TA_1$ and $TA_2$ is stronger than that of the longitudinal acoustic branch LA. According to the long-wave limit approximation, the shear elastic constant of α-U may have strong temperature dependence.

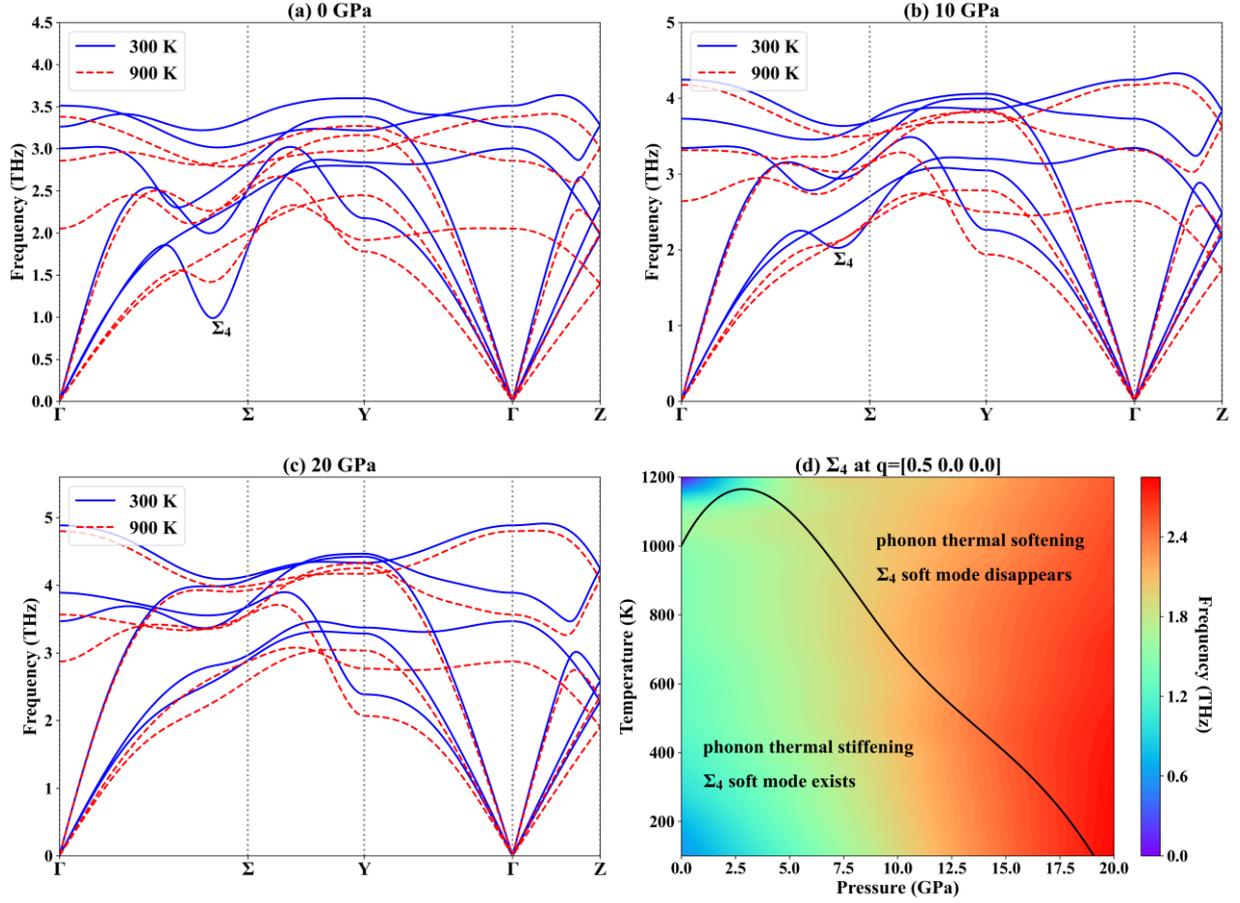

**Fig. 3**. (Color online) Phonon dispersion relations of α-U at 300 K and 900 K at different pressures: (a) 0 GPa; (b)10 GPa; (c) 20 GPa. (d) $\Sigma_4$ phonon frequency as a function of pressure and temperature at **q** = [0.5 0 0] for α-U.

At zero pressure, when the temperature reaches 900 K, the soft mode (Kohn anomaly) of $\Sigma_4$ is still clearly visible, which is consistent with the theoretical results of Bouchet *et al*.[7]. As the pressure increases, the soft-mode presence temperature range of $\Sigma_4$ decrease. For instance, at 10 GPa, the soft mode of $\Sigma_4$ is still visible at 300 K but it largely disappears at 900 K. The existence of the soft mode is completely invisible at 20 GPa, which is in accordance with the experimental measurements[6]. When the $\Sigma_4$ Kohn anomaly disappears, the overall frequency of $\Sigma_4$ decreases. This phenomenon may be attributed to the competition between electron and phonon coupling and phonon anharmonicity.

To discuss the phonon anharmonicity of α-U at high temperatures and pressures, we provide phonon spectral functions at different pressures at 900 K (Fig. 4). Although the phonon linewidth increases at high temperatures, the peak of the spectral function of α-U remains clear. Therefore, the phonon anharmonicity of α-U is weak. The weakening of the electron-phonon interaction may lead to the phonon thermal stiffening of $\Sigma_4$ at **q** = [0.5 0 0], and the Kohn anomaly disappears. To unravel the physical mechanism of the pressure- and temperature-dependent Kohn anomaly in α-U, it may be necessary to discuss its Fermi surface at high temperatures, such as V[57] and Nb[58].

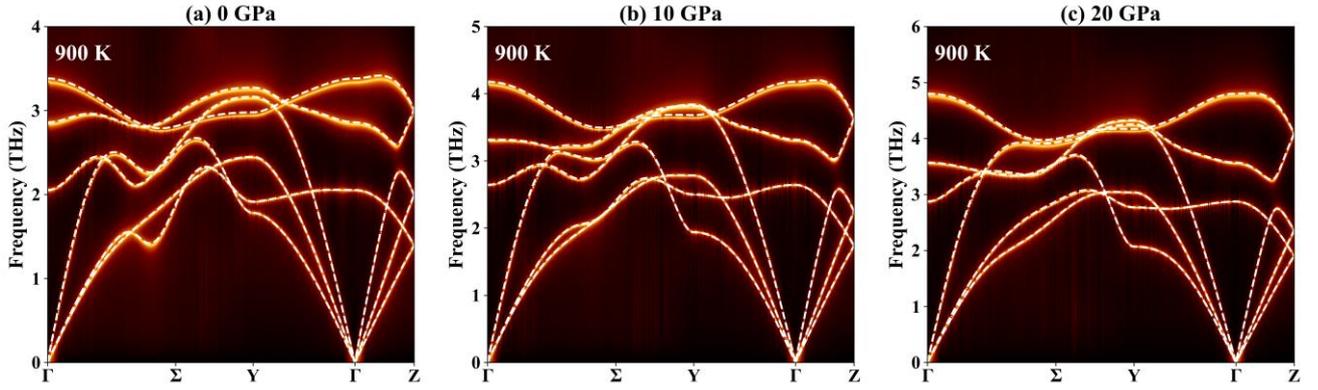

**Fig. 4**. (Color online) 900 K spectral function of α-U at different pressures. (a) 0 GPa; (b) 10 GPa; (c) 20 GPa. The white dashed line represents the phonon dispersion of the second-order TDEP.

### 3.3. Elastic properties at high temperatures and pressures

Furthermore, we used the MTP to predict the elastic properties of α-U at high temperatures and pressures. This is also the first systematic study on the variation in the α-U elastic constant with temperature and pressure. The single-crystal elastic constants of α-U at high temperature and pressure are shown in Fig. 5. According to the Born stability criterion, α-U exhibits mechanical stability. All the single-crystal elastic constants exhibited significant compression-hardening behavior. Under high pressure, $C_{12}$ and $C_{13}$ exhibited weak heating-induced hardening behavior, with a maximum change value not exceeding 10%. Additionally, $C_{23}$ is not affected by the temperature. Notably, $C_{44}$ ($C_{55}$, $C_{66}$) exhibited stronger temperature-dependent behavior compared to $C_{11}$ ($C_{22}$, $C_{33}$). For instance, at 100 GPa, between room temperature and 1000 K, $C_{44}$ ($C_{55}$, $C_{66}$) decreased by 14% (19%, 14%), whereas $C_{11}$ ($C_{22}$, $C_{33}$) decreased by 7% (3%, 6%). This is also consistent with the phonon thermal softening behavior of the acoustic modes of α-U near the Γ point.

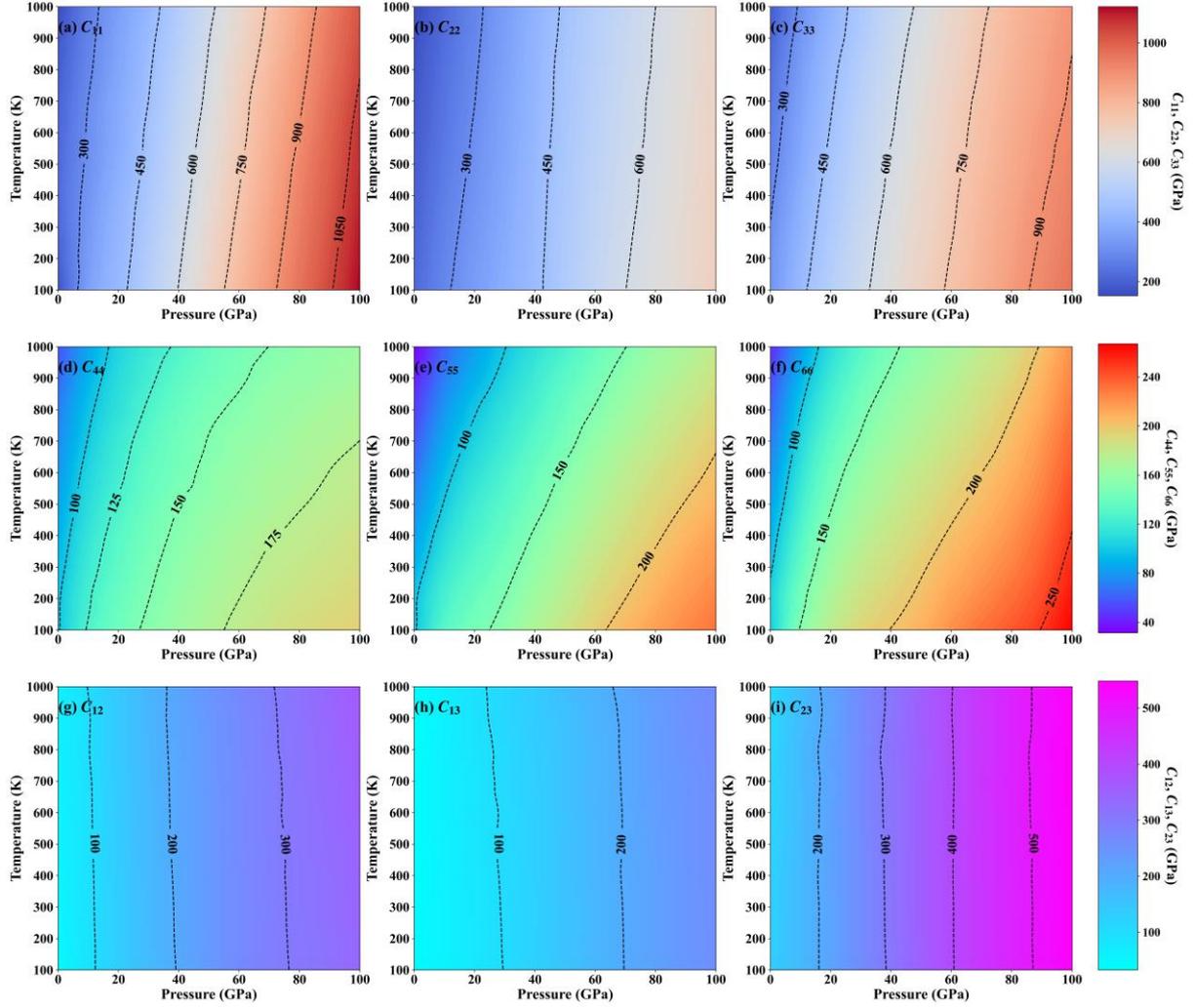

**Fig. 5**. (Color online) Single-crystal elastic constants of α-U as a function of pressure and temperature. (a) $C_{11}$; (b) $C_{22}$; (c) $C_{33}$; (d) $C_{44}$; (e) $C_{55}$; (f) $C_{66}$; (g) $C_{12}$; (h) $C_{13}$; (i) $C_{23}$.

The polycrystalline elastic modulus and Poisson's ratio of α-U were calculated using the single-crystal elastic constants obtained by the MTP. The results are presented in Fig. 6. At room temperature, the bulk (shear) modulus was 124.8 GPa (87.3 GPa), which is significantly higher than that of α-Pu[59]. Similar to the single-crystal elastic constants, the bulk modulus ($B$), Young's modulus ($E$), and shear modulus ($G$) exhibit significant compression hardening behavior. The Young's modulus and shear modulus have a stronger temperature dependence than the bulk modulus. At 0 GPa, $E$ and $G$ decrease by 46% and 50% between 300–1000 K, respectively. When the pressure is 100 GPa, $E$ and $G$ decrease by 13 and 15%, respectively. It is shown that temperature causes strong softening of the stiffness of α-U. The bulk modulus decreased by 21% (0 GPa) and 1.6% (100 GPa) between 300–1000 K, respectively. At room temperature, the Poisson's ratio of α-U is 0.21, which is similar to that of Pu (0.15– 0.21) and lower than that of most metals (generally greater than 0.3). The Poisson's ratio of α-U gradually increases with the increase in pressure and temperature. This indicates that the compressibility of α-U decrease at high temperatures and pressures. This is also consistent with the variation in Poisson's ratio for most metals, such as Al[60], β-Pu, and γ-Pu[59].

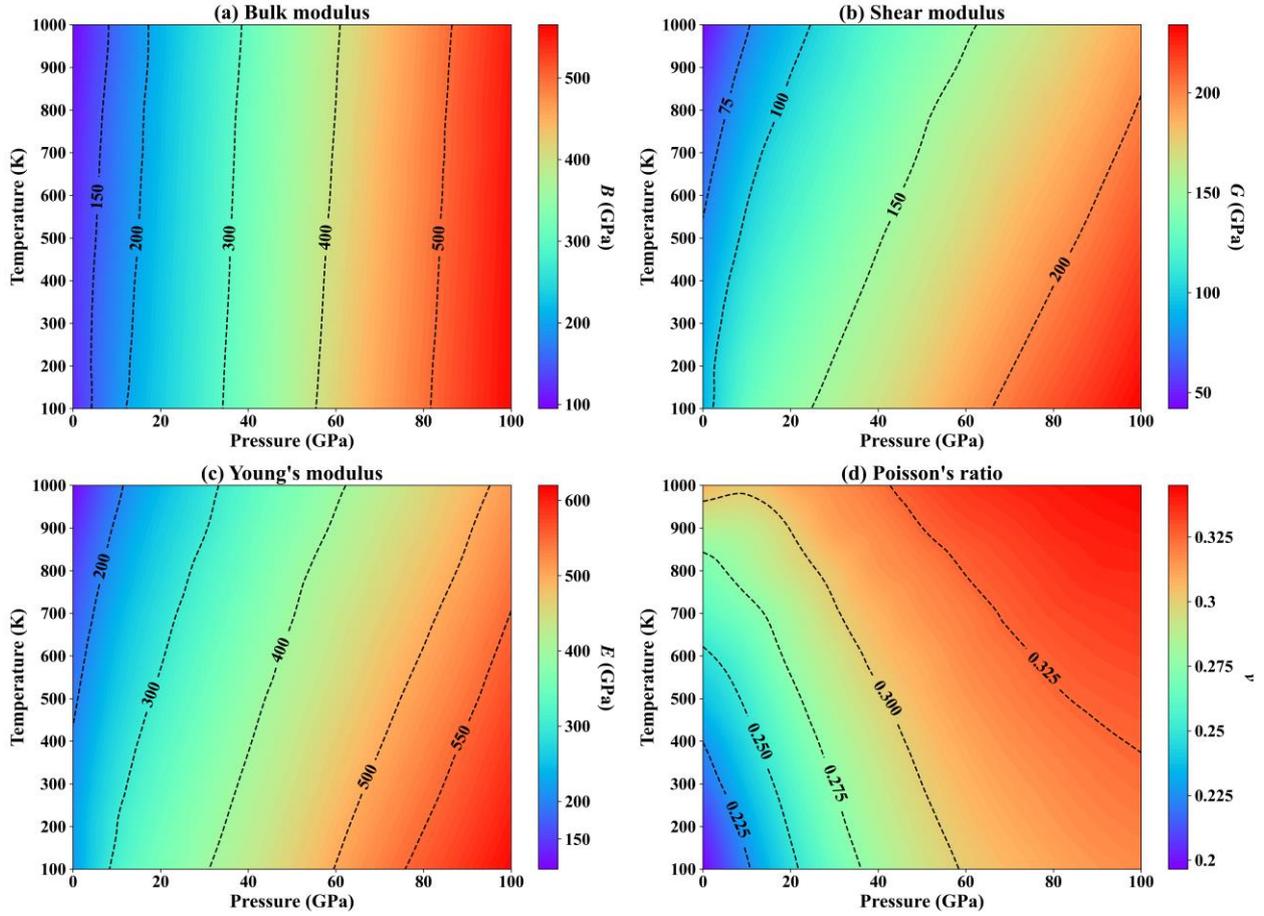

**Fig. 6**. (Color online) Bulk modulus ($B$), Young's modulus ($E$), shear modulus ($G$), and Poisson's ratio ($v$) of α-U as a function of temperature and pressure.

The polycrystalline sound velocity of α-U is shown in Fig. 7. Polycrystalline sound velocities increased with increasing pressure. Similar to the polycrystalline modulus behavior, the bulk sound velocity is less affected by temperature, whereas $C_L$ and $C_S$ exhibit a strong temperature dependence. As the temperature increases, $C_L$ and $C_S$ of α-U decrease rapidly. This study also provides a reference for future experimental measurements.

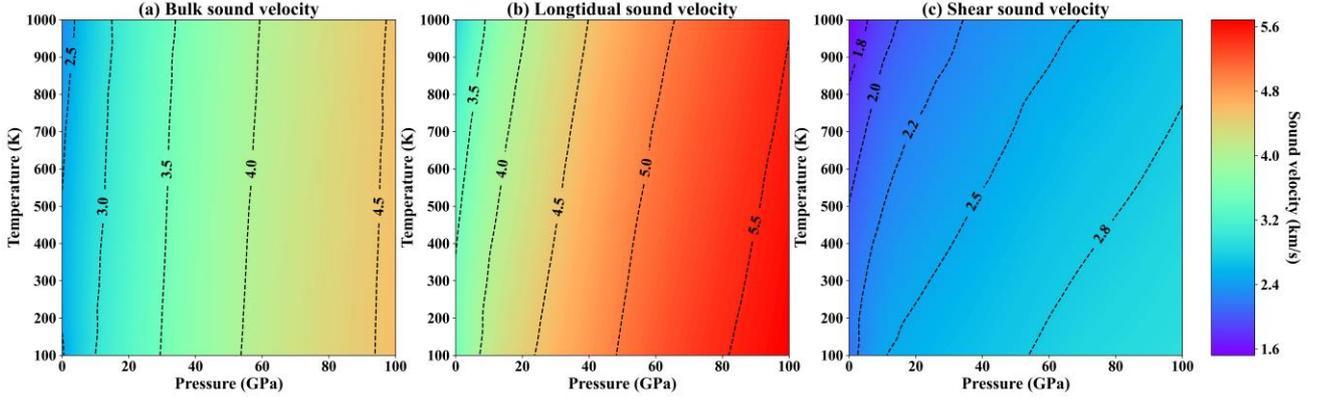

**Fig. 7**. (Color online) Polycrystalline sound velocity of α-U as a function of temperature and pressure. (a) Bulk sound velocity ($C_b$); (b) longitudinal sound velocity ($C_L$); (c) shear sound velocity ($C_S$).

### 3.4. Elastic anisotropy at high temperatures and pressures

Elastic anisotropy is a direct reflection of the spatial anisotropy of chemical bonds and it is related to properties such as the development of plastic deformation of crystals[61]. To measure the elastic anisotropy characteristics of α-U under high temperature and pressure, we obtained a universal anisotropy index ($A^U$). When the $A^U$ is larger than 0, the elastic anisotropy is stronger. At 300K, the $A^U$ of α-U is 0.34, which is higher than theoretical $A^U$ of α-Pu[62] (0.14) and α-Np (0.13)[63] and lower than fcc-Th[63] (1.35). This indicates that among the actinides, the elastic anisotropy of Th with cubic structure is higher than that of α-U (α-Np and α-Pu) with low symmetry. As shown in Fig. 8, α-U exhibits significant elastic anisotropy, and $A^U$ gradually increases with temperature and pressure. Notably, at 40-60 GPa, the $A^U$ changed very slowly and it was suppressed considerably. This suppression was also manifested in the DFT results at zero temperature[64].

To further measure the anisotropy of α-U, we calculated the Young's moduli in the [100], [010], and [001] directions and their ratios, as shown in Fig. 9. The growth rate of $E_{100}$ with pressure is significantly faster than that of $E_{010}$ and $E_{001}$, and the value was also much higher than that of the other two. At 300 K, when the pressure reached 100 GPa, $E_{100}$ increased by 738 GPa and reached 936 GPa. In contrast, $E_{010}$ ($E_{001}$) increases by 214 GPa (301.71 GPa), reaching 363 GPa (528.26 GPa). This indicates that the ability to resist compressive deformation in the [100] direction is significantly stronger than that in the other directions. The ratios of $E_{100}$ to $E_{010}$ and $E_{001}$ increased with pressure and temperature, whereas $E_{010}$:$E_{001}$ was hardly affected by temperature and pressure. The above results indicate that the anisotropy of α-U further increases at high temperatures and pressures. The rapid increase in $E_{100}$ with increasing pressure may be one reason for the enhanced anisotropy.

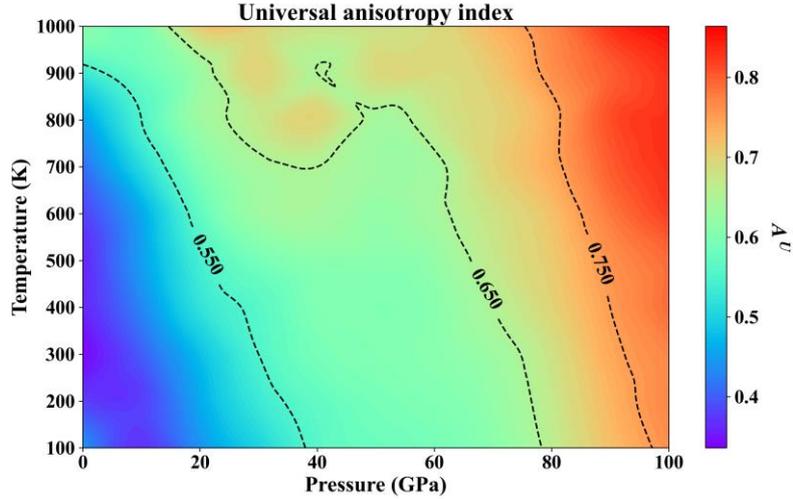

**Fig. 8**. (Color online) University anisotropy index ($A^U$) of α-U as a function of pressure and temperature.

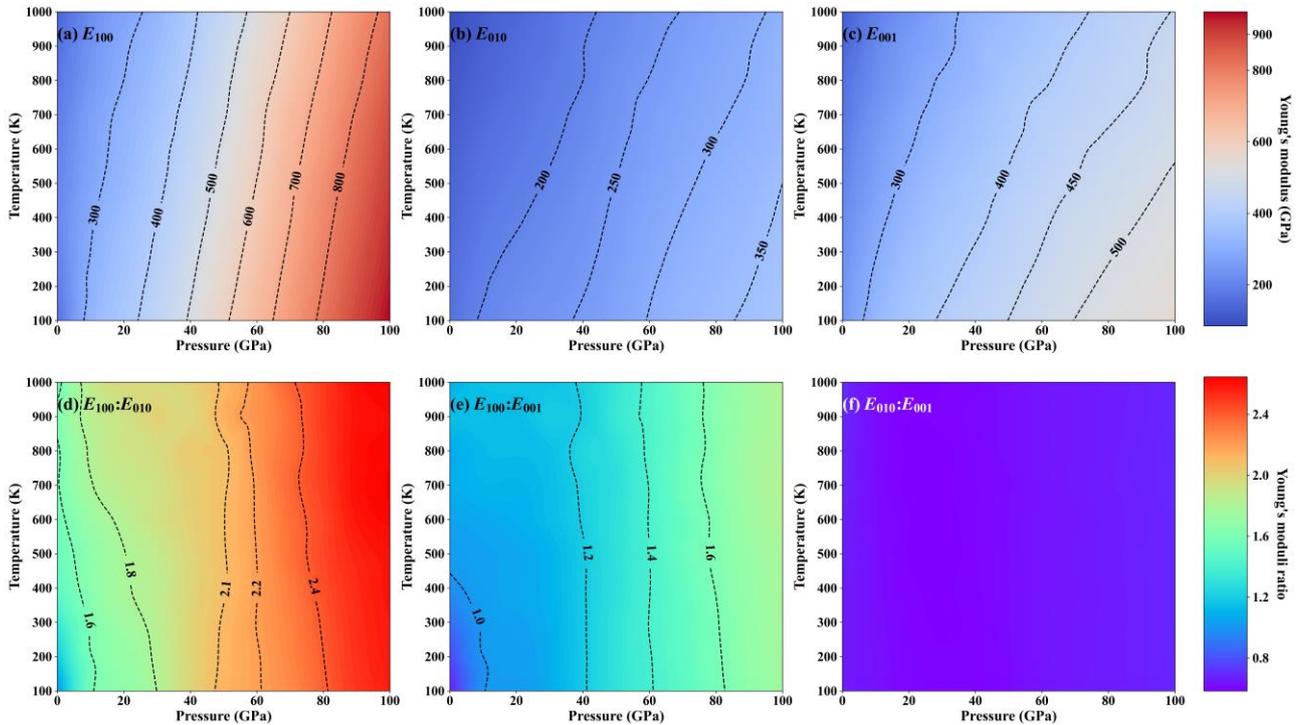

**Fig. 9**. (Color online) Young's modulus at different direction and Young's moduli ratio of the single-crystal α-U as a function of temperature and pressure. (a) $E_{100}$; (b) $E_{010}$; (c) $E_{010}$; (d) $E_{100}$:$E_{010}$; (e) $E_{100}$:$E_{001}$; (f) $E_{010}$:$E_{001}$.

## 4. Conclusions

In summary, we constructed a machine learning force field for α-U using a moment tensor potential method. MTP successfully reproduces the AIMD computational accuracy and has a higher computational efficiency. Based on the MTP, we predicted the lattice dynamics and elastic properties of α-U at high temperatures and pressures. Through a high-temperature lattice dynamics study, we observed that $\Sigma_4$ exhibits a Kohn anomaly and phonon thermal stiffening behavior within a certain temperature and pressure range. Based on phonon spectral function analysis, the phonon anharmonic effect of α-U has little effect on the phonon band structure. The variation in the Kohn anomaly may be related to the weakening of the electron-phonon coupling effect.

We studied the single-crystal elastic constant, polycrystalline modulus, and polycrystalline sound velocity of α-U at high temperatures and pressures. All the elastic properties exhibited a significant compression-induced hardening effect. Compared with other elastic constants, $C_{44}$, $C_{55}$, and $C_{66}$ exhibited significant heating-induced softening effects. Compared with $B$ ($C_b$), $E$ ($C_L$) and $G$ ($C_S$) decrease rapidly with increasing temperature. The Poisson's ratio shows that α-U is more difficult to compress under extreme conditions. By analyzing the direction-dependent Young's modulus and universal anisotropy index, we predict that α-U has a stronger anisotropic character at high temperatures and pressures.

In this study, the machine learning force field showed great promise for describing the physical properties of materials at high temperatures and pressures. Notably, the MTP obtained in this study is currently only applicable to α-U with an orthogonal structure, and it cannot correctly express the physical properties of the body-centered cubic γ-U structure. In the future, we will further improve the MTP to describe the physical properties of all uranium phases.

## Acknowledgments

This work was supported by the NSAF under Grant Nos. U1730248 and U1830101, the National Natural Science Foundation of China under Grant Nos. 11672274 and 12074214 and 11904282. The simulation was performed on resources provided by the Center for Comput. Mater. Sci. (CCMS) at Tohoku University, Japan.

# Supplementary material: Lattice dynamics and elastic properties of α-U at high-temperature and high-pressure by machine learning potential simulations

Hao Wang[1, 2], Xiao-Long Pan[1, 2], Y. F. Wang[2], Xiang-Rong Chen[1*], Yi-Xian Wang[3], Hua-Yun Geng[2, 4*]

[1] College of Physics, Sichuan University, Chengdu 610065, People's Republic of China

[2] National Key Laboratory of Shock Wave and Detonation Physics, Institute of Fluid Physics, CAEP, Mianyang 621900, People's Republic of China

[3] College of Science, Xi'an University of Science and Technology, Xi'an 710054, People's Republic of China

[4] HEDPS, Center for Applied Physics and Technology, and College of Engineering, Peking University, Beijing 100871, People's Republic of China

## A. Structure properties at high temperatures and pressures

Based on the NPT molecular dynamics simulation, the relationship between the lattice parameters of α-U and the temperature and pressure is obtained. Figure S1 shows the variation trend of the structural parameters of α-U under high temperature and high pressure. Figures S1(a) and S1(b) show the volume thermal expansion and linear thermal expansion coefficients of α-U at different pressures. At different pressures, MTP results are the same as AIMD. At zero pressure, the theoretically calculated results agree with the experimental data[1, 2]. With the increase of pressure, the volume thermal expansion of α-U is significantly suppressed. Similar to the behavior at normal pressure, the *b*-axis still exhibits negative linear thermal expansion behavior at high pressure (Fig. S1(b)). The relative volume as a function of pressure and axial ratios as a function of the relative volume are given in Fig. S1(c) and Fig. S1(d). The theoretical results of MTP agree well with the experiments[3]. The *c*/*a* ratio is independent of temperature. This is mainly because *a* and *c* have similar linear thermal expansion behavior.

* *Corresponding authors. E-mail:* xrchen@scu.edu.cn, s102genghy@caep.cn

The negative thermal expansion behavior of *b* is the reason for the decrease of *b*/*a* and *b*/*c* with temperature.

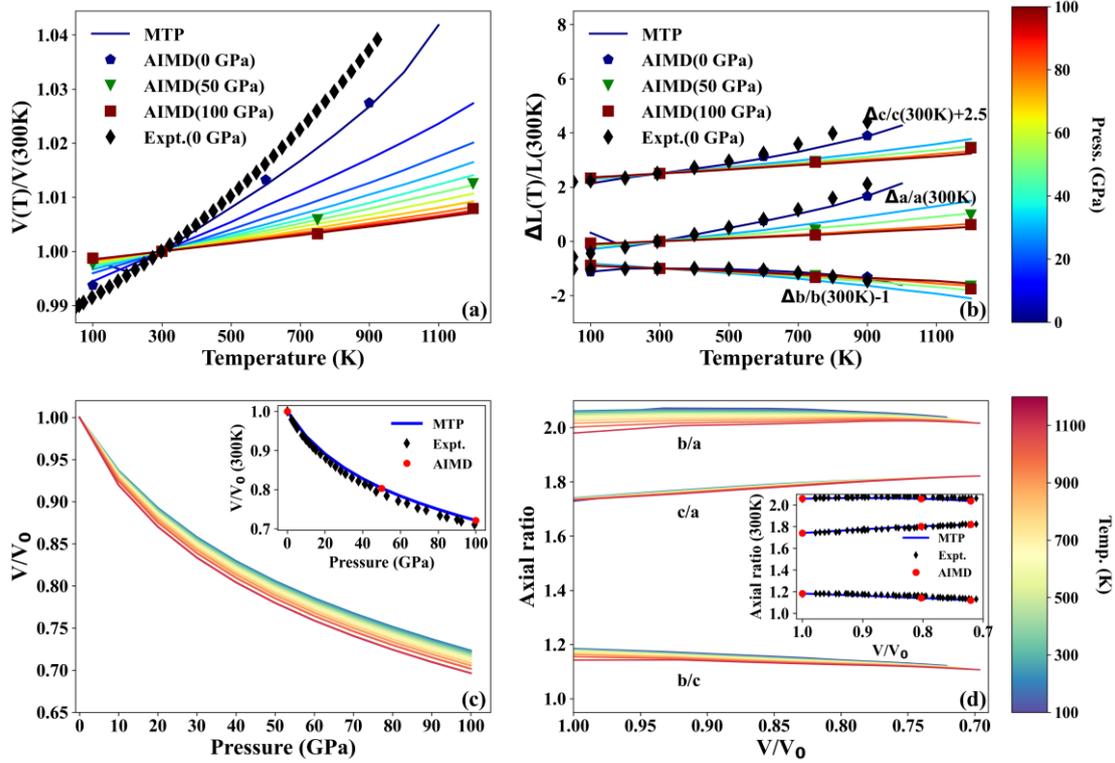

**Fig. S1**. (Color online) (a) The volume thermal expansion of α-U at different pressure. (b) The linear thermal expansion of α-U at different pressure. (c) Pressure-volume curves at different temperatures. (d) Axial ratios as a function of the relative volume at different temperatures. The inset is a comparison with experimental values at room temperature. The solid line represents the MTP results. The solid black diamonds represent the experimental results[1-3], and the solid patterns in other colors represent the AIMD results.

**B. The elastic constants by using AIMD, MTP, EAM, MEAM, and ADP at 0 GPa**

Figure S2-S5 shows the variation of the elastic properties of α-U with temperature at zero pressure from MTP, AIMD and classical potentials simulations. The variation trends of the elastic constants with temperature calculated by AIMD and MTP are consistent with the experimental results[2, 4]. Compared with the severe deviation of EAM from the experiment, MEAM performs better.

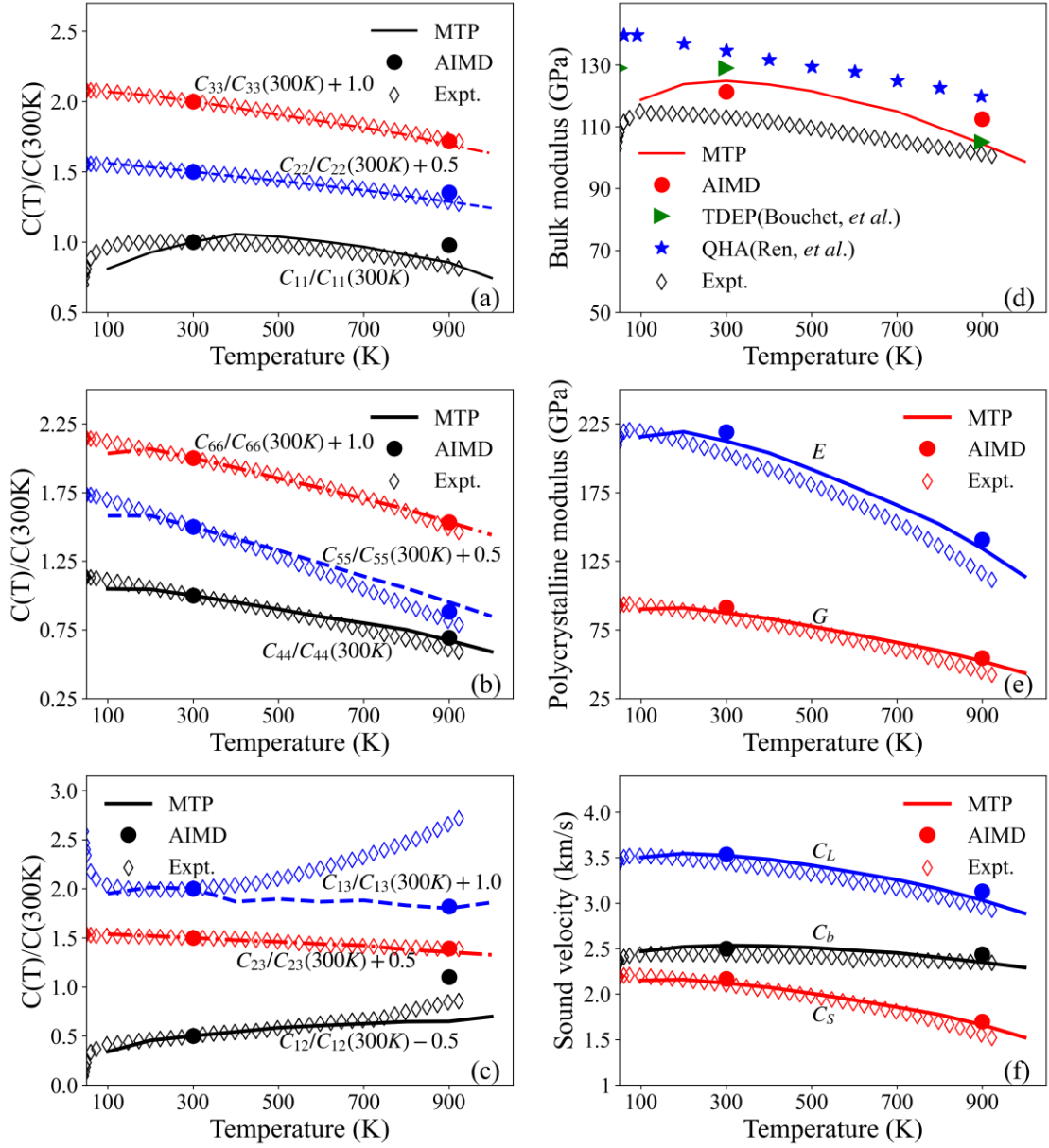

**Fig. S2**. (Color online) Elastic properties of α-U at 0 GPa as a function of temperature. (a) $C_{11}$, $C_{22}$, $C_{33}$; (a) $C_{44}$, $C_{55}$, $C_{66}$; (c) $C_{12}$, $C_{13}$, $C_{23}$; (b) Bulk modulus (*B*); (e) Young's modulus (*E*) and Shear modulus (*G*); (f) Bulk ($C_b$), longitudinal ($C_L$) and shear sound velocities ($C_S$). Lines represent MTP results, solid circle pattern represents AIMD results, the solid right triangle represents TDEP results[5], the solid five-pointed star pattern represents QHA results[6], and open black diamonds represent experimental values[2, 4].

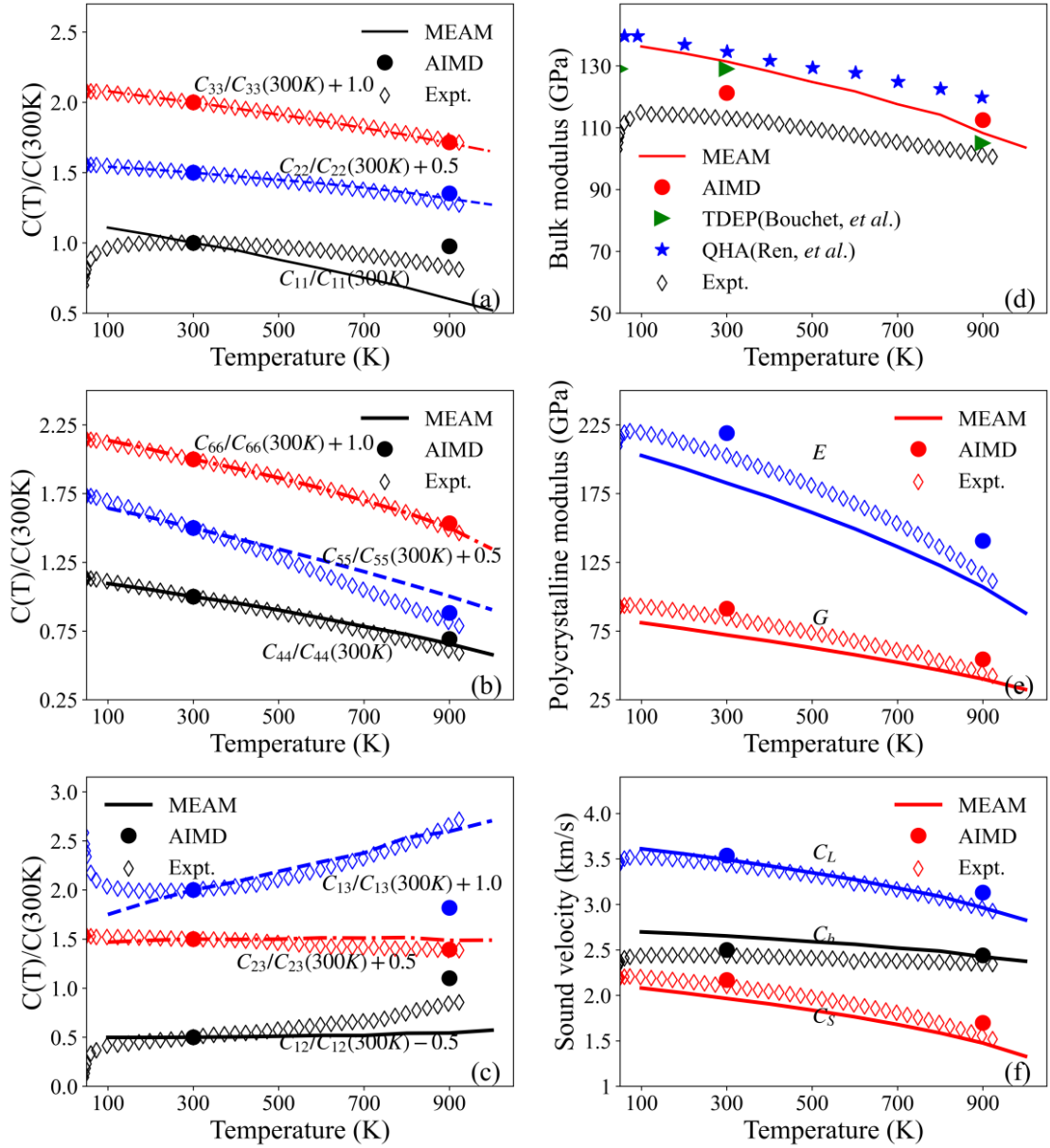

**Fig. S3**. (Color online) Elastic properties of α-U at 0 GPa as a function of temperature. (a) $C_{11}$, $C_{22}$, $C_{33}$; (a) $C_{44}$, $C_{55}$, $C_{66}$; (c) $C_{12}$, $C_{13}$, $C_{23}$; (b) Bulk modulus ($B$); (e) Young's modulus ($E$) and Shear modulus ($G$); (f) Bulk ($C_b$), longitudinal ($C_L$) and shear sound velocities ($C_S$). Lines represent MEAM[7] results, solid circle pattern represents AIMD results, the solid right triangle represents TDEP results[5], the solid five-pointed star pattern represents QHA results[6], and open black diamonds represent experimental values[2, 4].

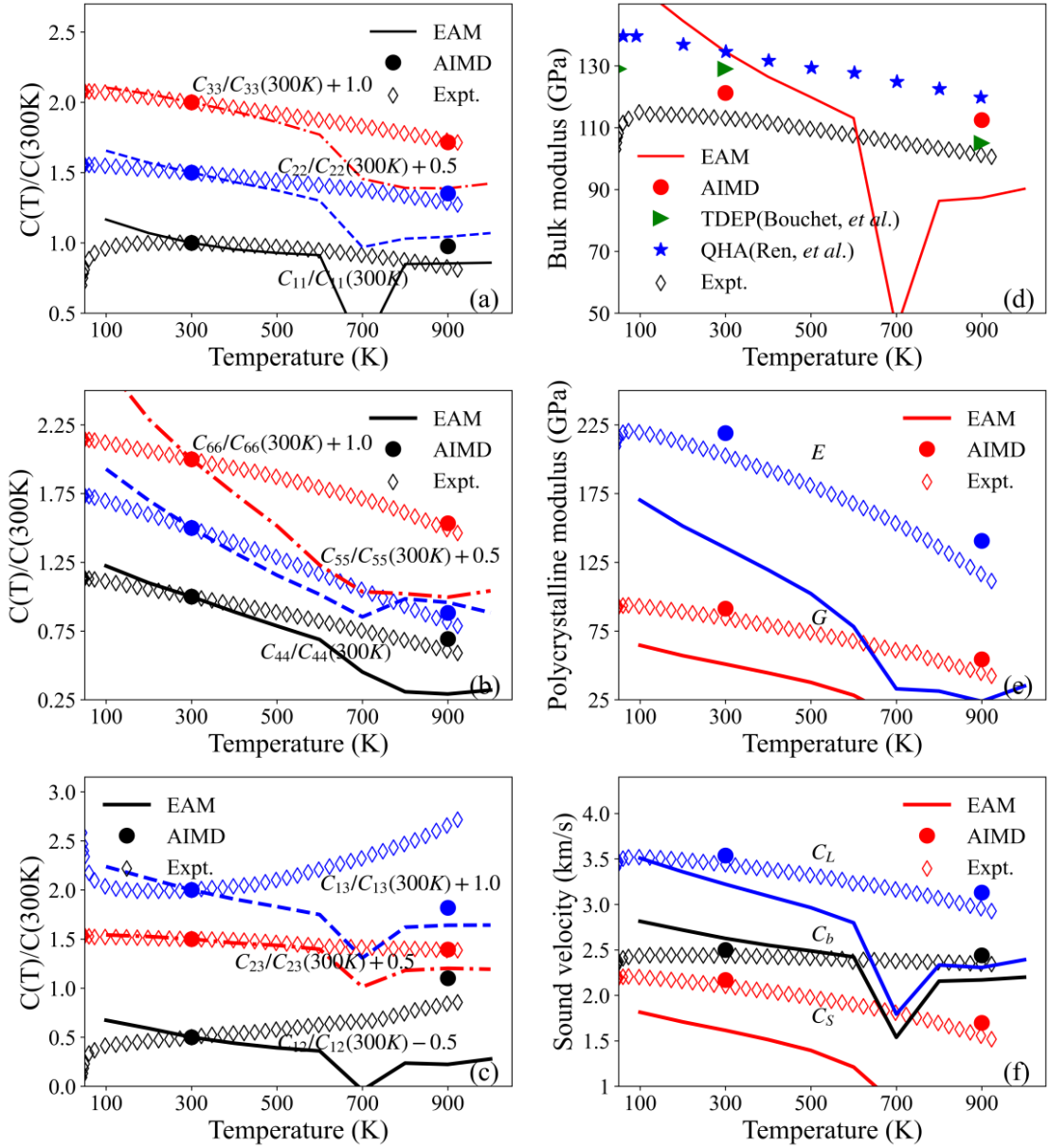

**Fig. S4**. (Color online) Elastic properties of α-U at 0 GPa as a function of temperature. (a) $C_{11}$, $C_{22}$, $C_{33}$; (a) $C_{44}$, $C_{55}$, $C_{66}$; (c) $C_{12}$, $C_{13}$, $C_{23}$; (b) Bulk modulus ($B$); (e) Young's modulus ($E$) and Shear modulus ($G$); (f) Bulk ($C_b$), longitudinal ($C_L$) and shear sound velocities ($C_S$). Lines represent EAM[8] results, solid circle pattern represents AIMD results, the solid right triangle represents TDEP results[5], the solid five-pointed star pattern represents QHA results[6], and open black diamonds represent experimental values[2, 4].

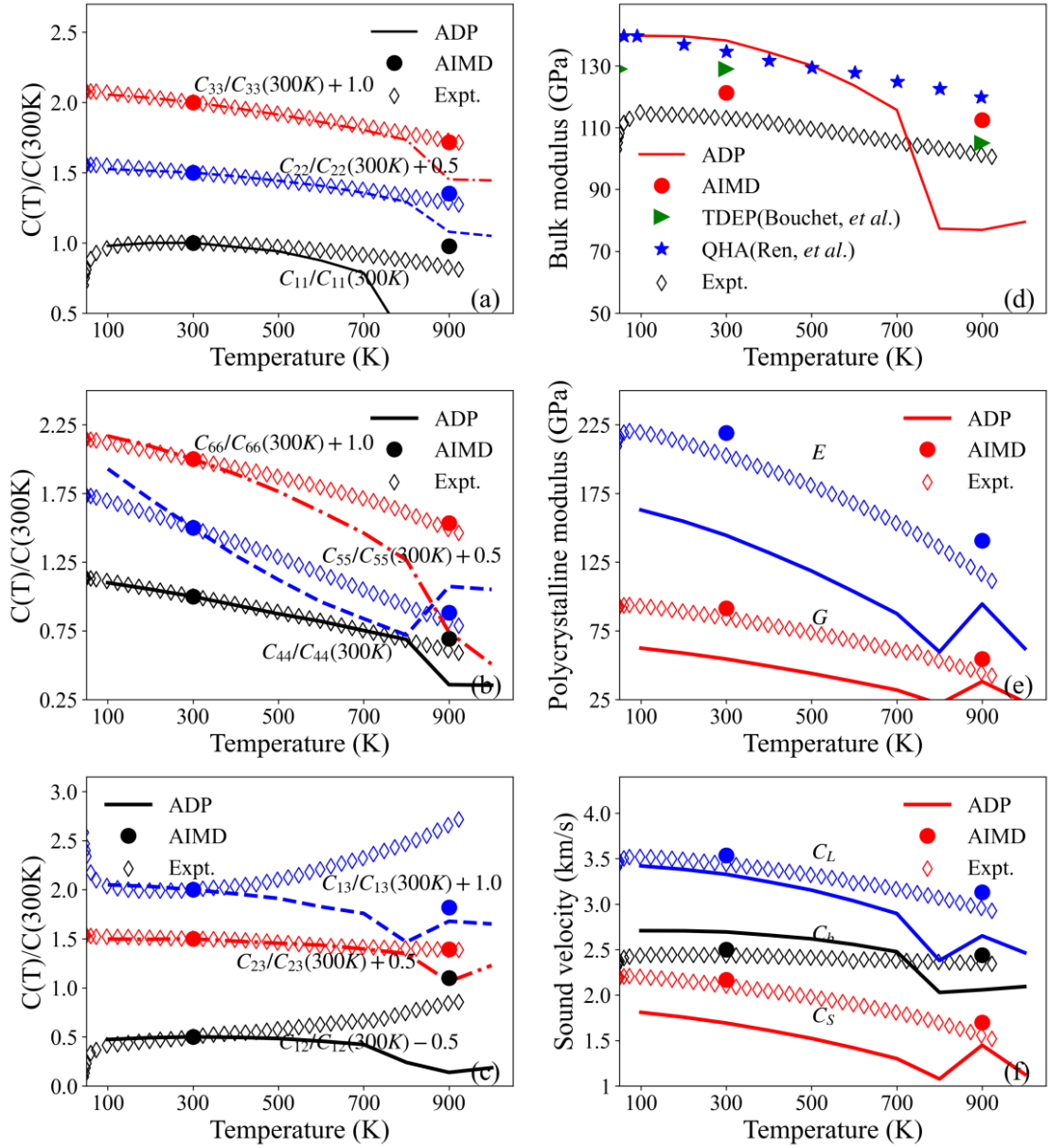

**Fig. S5**. (Color online) Elastic properties of α-U at 0 GPa as a function of temperature. (a) $C_{11}$, $C_{22}$, $C_{33}$; (a) $C_{44}$, $C_{55}$, $C_{66}$; (c) $C_{12}$, $C_{13}$, $C_{23}$; (b) Bulk modulus ($B$); (e) Young's modulus ($E$) and Shear modulus ($G$); (f) Bulk ($C_b$), longitudinal ($C_L$) and shear sound velocities ($C_S$). Lines represent ADP[9] results, solid circle pattern represents AIMD results, the solid right triangle represents TDEP results[5], the solid five-pointed star pattern represents QHA results[6], and open black diamonds represent experimental values[2, 4].

## C. The spin-orbit interaction effect

To better illustrate the effect of spin-orbit interactions, we compare the effects on elastic constants and phonon spectra with and without spin-orbit interactions. From the results, the spin-orbit interactions have little effect on the elastic constants. Although the spin-orbit interaction has an effect on the $\Sigma_4$ optical branch of the phonon spectrum, this effect does not affect our conclusions. And our potentials also show consistency with the experiments in Fig. 2.

**Table RI.** The calculated lattice constant and elastic constants of α-U under 0 GPa. In addition, the others' theoretical results in 0 K and experiment values in 300 K are listed as comparison.

|  | PAW | PAW+SOC | PAW+SOC[10] | PAW[11] | FP-LMTO[12] | Expt.[4, 13] |
|---|---|---|---|---|---|---|
| $a$ (Å) | 2.833 | 2.834 | 2.797 | 2.810 | 2.845 | 2.854 |
| $b$ (Å) | 5.828 | 5.832 | 5.867 | 5.861 | 5.818 | 5.869 |
| $c$ (Å) | 4.931 | 4.928 | 4.893 | 4.918 | 4.996 | 4.955 |
| $V$ (Å³) | 20.36 | 20.36 | 20.07 | 20.50 | 20.67 | 20.75 |
| $C_{11}$ (GPa) | 296 | 291 | 296 | 295 | 300 | 215 |
| $C_{22}$ (GPa) | 221 | 223 | 216 | 215 | 220 | 199 |
| $C_{33}$ (GPa) | 349 | 332 | 367 | 347 | 320 | 267 |
| $C_{44}$ (GPa) | 153 | 152 | 153 | 143 | 150 | 124 |
| $C_{55}$ (GPa) | 125 | 119 | 129 | 130 | 93 | 73 |
| $C_{66}$ (GPa) | 102 | 99 | 99 | 102 | 120 | 74 |
| $C_{12}$ (GPa) | 61 | 59 | 60 | 68 | 50 | 46 |
| $C_{13}$ (GPa) | 27 | 27 | 29 | 25 | 5 | 22 |
| $C_{23}$ (GPa) | 147 | 141 | 141 | 149 | 110 | 108 |
| $B$ (GPa) | 146 | 143 | 149 | 147 | 129 | 113 |
| $G$ (GPa) | 111 | 109 | 108 | 108 | 112 | 84.4 |
| $E$ (GPa) | 266 | 260 | 261 | 261 | 262 | 202.6 |

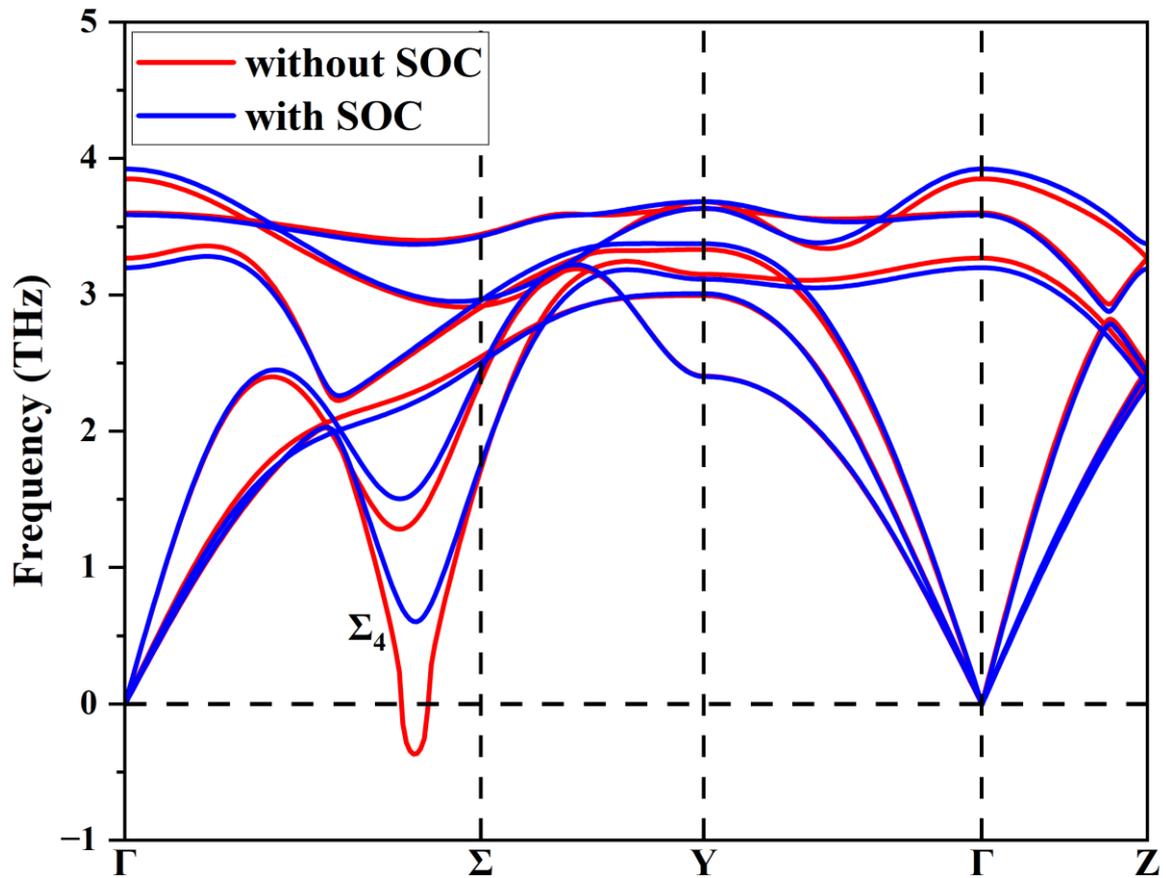

**Fig. S6.** (Color online) Phonon dispersion obtained with and without SOC.